# Imaging the Flow of Holes from a Collimating Contact in Graphene

Sagar Bhandari,[1,2] Mary Kreidel,[1] Alexander Kelser,[3] Gil-Ho Lee,[3,4] Kenji Watanabe,[5] Takashi Taniguchi,[5] Philip Kim[1,3] and Robert M. Westervelt[1,3*]

[1]School of Engineering and Applied Sciences, Harvard University, Cambridge, MA 02138, USA

[2]Department of Physics and Engineering, Slippery Rock University, Slippery Rock PA 16057, USA

[3]Department of Physics, Harvard University, Cambridge, MA 02138, USA

[4]Department of Physics, Pohang University of Science and Technology, Pohang, Republic of Korea

[5]National Institute for Materials Science, 1-1 Namiki, Tsukuba, 305-0044, Japan

## Abstract

A beam of holes formed in graphene by a collimating contact is imaged using a liquid-He cooled scanning probe microscope (SPM). The mean free path of holes is greater than the device dimensions. A zigzag shaped pattern on both sides of the collimating contact absorb holes that enter at large angles. The image charge beneath the SPM tip defects holes, and the pattern of flow is imaged by displaying the change in conductance between contacts on opposite sides, as the tip is raster scanned across the sample. Collimation is confirmed by bending hole trajectories away from the receiving contact with an applied magnetic field. The SPM images agree well with ray-tracing simulations.

**Physics Subject Headings** in the Discipline of Condensed Matter & Materials Physics: Physical-systems / Two-dimensional-systems / Graphene; Transport phenomena / Ballistic transport; Techniques / Scanning-techniques / Scanning-probe-microscopy.

Graphene displays remarkable electronic properties, including ballistic electron transport, Klein tunneling of electrons through potential barriers, and the anomalous quantum Hall effect [1-8]. Monolayer graphene samples encapsulated in hexagonal boron nitride (hBN) sheets achieve unusually high carrier mobility by minimizing electron scattering [9]. Electrons in hBN-graphene-hBN layered structures travel along ballistic trajectories over distances comparable to the size of the device that enable ballistic electronics based on the manipulation of electron beams [1,10], making it possible to achieve novel Dirac fermionic optics, such as negative refraction [11] and Veselago lensing [12].

Imaging the flow of electrons, or holes, provides a direct way to understand ballistic flow in graphene. A cooled scanning probe microscope (SPM) can image the flow of electrons through a two-dimensional electron gas (2DEG) by displaying the change in conductance between two point-contacts as the tip is raster scanned above the sample [13,14]. The tip creates an image charge in the 2DEG that deflects charge carriers and removes them from the pattern of flow. Displaying the conductance change *vs.* tip position provides an image of the carrier flow. Scanned probe imaging has proven to be a useful approach for understanding electron motion [15-18]. In previous work on graphene, we imaged the pattern of electron flow from a collimating contact [10] and the cyclotron orbits of electrons in the magnetic focusing regime [19,20].

In the present study, we use a cooled SPM to image the flow of holes in graphene from a collimating contact. Because the band structure of graphene is symmetric about the Dirac point for small densities, the hole mobility is expected to be similar to the electron mobility [21,22]. The pattern of flow between two contacts on opposite sides of the sample is imaged by displaying the change in conductance as the SPM tip is raster scanned across the graphene. Confirmation that the collimating contact forms a hole beam is provided by applying a perpendicular magnetic field $B$ that bends hole trajectories away from the receiving contact. Creating complimentary electron [10] and hole beams opens new approaches to ballistic graphene devices.

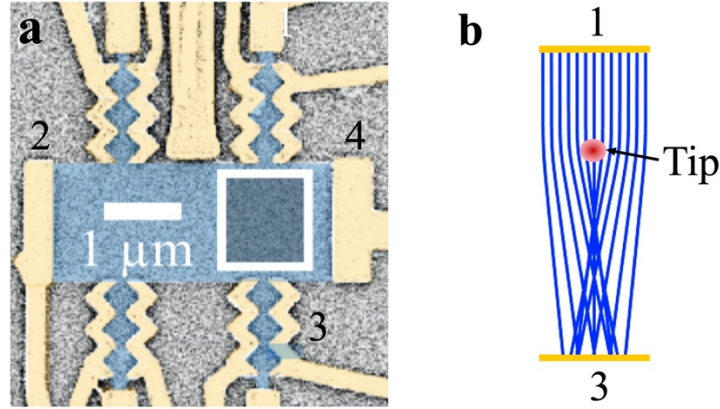

FIG. 1: (a) Scanning electron micrograph of the graphene device. Four zigzag shaped collimating contacts are positioned along the top and bottom sides of the device, and a large contact covers each end. The white square indicates the area imaged by the scanned probe microscope (SPM). (b) Ray-tracing simulation of hole trajectories travelling from contact 1 to contact 3, deflected by the image charge beneath the SPM tip (red dot).

A scanning electron micrograph of the collimating contact device is shown in Fig. 1(a). An exfoliated monolayer graphene sheet was encapsulated between two layers of mechanically cleaved hexagonal boron nitride (hBN) to enhance the carrier mobility. The device sits on a heavily doped Si substrate topped by a 285 nm thick layer of $SiO_2$, which acts as a back gate. Using a dry transfer technique, the bottom hBN, graphene, and top hBN flakes were stacked onto the substrate. Using reactive-ion etching, the device was shaped into a Hall bar that has two wide end contacts and two collimating contacts on either side. Each collimating contact consists of a narrow contact that emits electrons or holes into the graphene and two zigzag contacts, one on either side, that absorb carriers entering at large angles. Collimation can be turned off by floating the zigzag side contacts. For electrons, the half angle of the collimated beam is 9°, measured using our cooled SPM [10]. The rectangular graphene channel has dimensions 1.6 x 5.0 µm² and the collimating contacts on either side are separated by 1.6 µm. To make high quality contacts, chromium and gold layers were evaporated onto the freshly etched graphene edge immediately after etching [23].

The device was mounted inside our SPM and cooled to the temperature 4.2 K. For this geometry, the back-gate capacitance is $C_G$ = 30 fF. The hole density is $p = C_G(V_D - V_G)/e$, where $V_G$ is the back-gate voltage, $V_D$ is the back-gate voltage that puts the Fermi level at the Dirac point, and $e$ is the fundamental charge. The transmission $T$ of holes between collimating contact 1 and contact 3 was measured by passing a current between contact 1 and contact 2 and measuring the voltage difference $V_s$ between contact 3 and contact 4. The accumulation of holes at contact 3 raises the potential of contact 3. The transmission $T_m$ of holes from contact 1 to contact 3 is proportional to the measured transresistance $R_m = (V_s/I_i)$.

A cooled scanning probe microscope (SPM) is used to image the motion of holes through the graphene sample. The technique is adapted from previous imaging experiments for two-dimensional electron gases in graphene and GaAs/AlGaAs heterostructures [10,13-14,16,19-20]. Holes are emitted from the top collimating hole contact (contact 1) and travel along ballistic trajectories to the bottom contact 3, in Fig. 1(a). The zigzag sides on contact 3 are floated to turn off collimation, so holes can enter over a wide range of angles. To image the ballistic transmission of holes from contact 1 to contact 3, the silicon SPM tip is held above the encapsulated graphene creating an image charge below. As shown in Fig. 1(b), the local increase in hole density beneath the tip acts as a lens that focuses and deflects ballistic hole trajectories. An image of ballistic hole flow can be obtained by displaying the change $\Delta R_m$ as the SPM tip is raster scanned above the sample, as shown in the data below.

Working in the ballistic limit, we use ray tracing to simulate hole trajectories in graphene under the influence of a magnetic field $B$. The image charge forms a peak in hole density $\Delta p_{tip}$ in the graphene sheet directly below the SPM tip:

$$\Delta p_{tip} = - qd/2\pi e\varepsilon(a^2 + d^2)^{3/2} \tag{1}$$

where the tip is modeled as a point charge $q$ at height $d = 70$ nm above the graphene sheet, $a$ is the radial distance in the sheet away from the tip position, $e$ is the fundamental charge, and $\varepsilon$ is the dielectric constant of hBN.

We obtain the force that the image charge exerts on a ballistic hole traveling nearby by balancing the flow away from the tip caused by the peak in the Fermi energy $E_F$, with the flow toward the tip caused by the dip in potential energy $U$ that attracts holes to the tip. The total chemical potential $E_F(\mathbf{r}) + U(\mathbf{r})$ of the hole gas is constant, where $\mathbf{r}$ is the position. The tip-modified Fermi energy is:

$$E_F(\mathbf{r}) = hv_F[(p_i(\mathbf{r}) + \Delta p_{tip}(\mathbf{r}))/4\pi]^{1/2}, \tag{2}$$

where $h$ is Planck's constant, and $p_i$ is the hole density without the tip present. Because the chemical potential is constant, the force on a hole from the tip is $-\nabla U(\mathbf{r}) = \nabla E_F(\mathbf{r})$. Using the dynamical mass of holes in graphene $m^* = h(p/\pi)^{1/2}/2v_F$, where $v_F$ is the speed associated with the conical energy bands, we find the equation of motion:

$$d^2\mathbf{r}/dt^2 = (v_F^2/2p) \nabla p(\mathbf{r}). \tag{3}$$

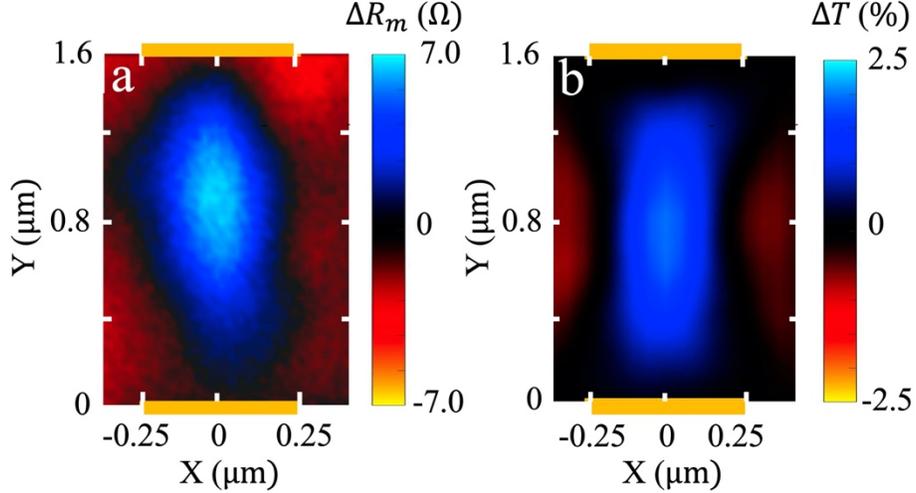

FIG. 2: (a) SPM image of hole flow between contact 1 (top orange bar) and contact 3 (bottom orange bar) in the absence of a magnetic field. The measured change in transresistance $\Delta R_m$ is displayed vs. tip position. (b) Simulated change in hole transmission $\Delta T$ between contacts 1 and 3 vs. tip position, predicted by the ray-tracing model.

The SPM tip creates a force that pulls holes beneath the tip. Under a magnetic field, the Lorentz force $\mathbf{F} = e\mathbf{v} \times \mathbf{B}$ acts on a hole with group velocity $\mathbf{v}$.

Figure 1(b) is a ray tracing illustration that shows how the image charge beneath the SPM tip changes the transmission $T$ by bending hole trajectories that travel nearby. The image charge acts as a lens that can focus hole paths onto the receiving contact. This behavior is in sharp contrast to what happens for electrons, where the tip potential push es the carriers away and defocuses their flow toward the receiving contact [20]. Simulated SPM images of hole flow from the collimating contact 1 to the receiving contact 3, shown in Figs. 2(b) and 3(b) below, are made by computing the change in transmission $\Delta T$ between the two contacts as the SPM tip is raster scanned across the area between them. The zigzag sides of contact 3 are floated to turn off collimation, so holes can enter over a wide range of angles.

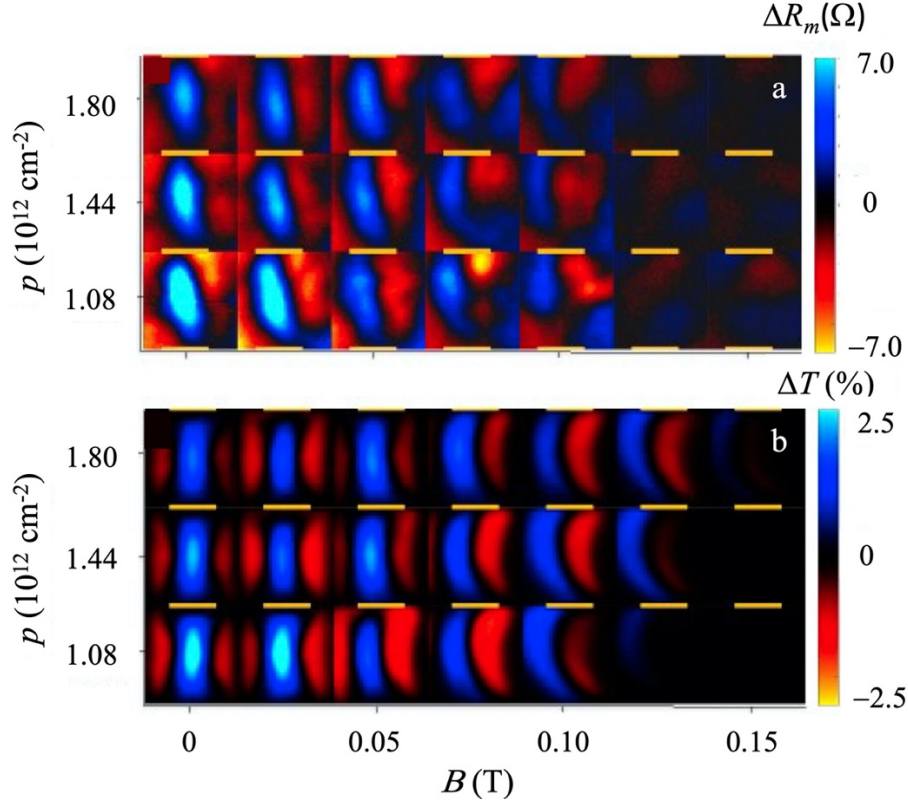

FIG. 3: A set of (a) SPM experimental and (b) simulated images of hole flow in graphene from collimating contact 1 to receiving contact 3 vs. hole density $p$ and magnetic field $B$; here $\Delta R_m$ is the change in the measured transresistance, and $\Delta T$ is the simulated change in transmission between contacts 1 and 3. As $B$ is increased, the hole paths curve away from contact 3, causing the signal to disappear. In blue regions, the tip focuses holes into contact 3, but in red regions, the tip deflects holes away from contact 3.

Figure 2(a) shows an SPM image of hole flow from the top collimating contact (contact 1) of the device to the bottom contact (contact 3) at 4.2 K with no magnetic field applied. When the tip is in the center of the channel, the measured change in transresistance $\Delta R_m$ is positive, because the flow of holes is focused into the receiving contact by the lens beneath the tip formed by the image charge. When the tip is away from the center, it bends hole trajectories away from contact 3, reducing the transmission. Simulations of the change in transmission $\Delta T$ displayed in Fig. 2(b), agree well with the experimental images.

We studied the degree of collimation by applying a perpendicular magnetic field $B$. The applied magnetic field bends the hole trajectories along cyclotron orbits that eventually curve away from the receiving contact and result in a reduction of flow. If a collimated hole beam with a small spread angle is emitted, the received signal $\Delta R_m$ will fall away rapidly as $B$ is increased, but if holes enter over a wide angle, the applied magnetic field will have less effect.

Figure 3(a) shows a series of SPM images of hole flow between contacts 1 and 3 taken at hole densities ranging from $p = 1.08\times10^{12}$ cm$^{-2}$ to $1.80\times10^{12}$ cm$^{-2}$ and magnetic fields $B = 0$ T to 0.15 T. These images are in good agreement with the simulations shown in Fig. 3(b). The blue regions in the center of the images in Figs. 2 and 3 show that the image charge beneath the tip acts to focus holes into the receiving contact, as illustrated in Fig. 1(b). The imaging signal is strongest at lower hole densities, where the image charge induced by the tip is a greater fraction of the original hole density.

As the magnetic field $B$ is increased in Fig. 3(a), the hole paths bend away from the bottom contact, and the strength of the imaging signal decreases until it disappears. The curvature results from the Lorentz force, and the counterclockwise bend affirms that the carriers are positively charged, i.e. they are holes. The signal eventually disappears at $B = 0.15$ T, when the Lorentz force is strong enough to bend the hole beam entirely away from contact 3. The curvature of the hole paths in Fig. 3(a) is greatest at lower hole densities $p$, in agreement with the expression for the cyclotron orbit $d_c = h(p/\pi)^{1/2}/eB$.

To summarize, images of carrier flow taken by our cooled SPM show that a beam of holes is emitted into graphene by the collimating contact shown in Fig. 1(a). In addition, we find that the image charge beneath the SPM tip can act as a focusing lens for holes. These results compliment our previous demonstration of a collimating contact for electrons in graphene [10], where the tip potential deflects electrons and defocuses the electron beam. The SPM images are in good agreement with ray-tracing simulations for experimentally relevant carrier densities and magnetic fields. The ability to make complimentary beams of electrons and holes paves the way for novel approaches to ballistic devices based on massless Dirac fermions in graphene.

**Acknowledgements**

S.B. conducted the imaging experiments, aided by M.K. and A.K., using a graphene device designed and fabricated by G.H.L. The investigators were advised by P.K. and R.M.W. Hexagonal boron nitride samples were provided by K.W. and T.T. Our SPM research was supported by the DOE Office of Basic Energy Sciences, Materials Sciences and Engineering Division, under grant DE-FG02-07ER46422. Graphene sample fabrication was supported by the ARO grant W911NF-17-1-0574. Analysis was supported in part by NSF DMR-1231319, and nanofabrication by NSF NNCI site award ECCS-1541959. The growth of hexagonal boron nitride crystals was supported in Japan by the Elemental Strategy Initiative conducted by the Ministry of Education, Culture, Sports, Science and Technology, Japan, and Creating the Seeds for New Technology by Award JPMJCR15F3, Japan Science and Technology Agency.